# Clusters in Very High Energy Cosmic Rays

A.A. Mikhailov

Yu.G. Shafer Institute of Cosmophysical Research and
Aeronomy, 31 Lenin Ave., 677891 Yakutsk, Russia

## Abstract

Arrival directions of cosmic rays with the energy $E > 4 \cdot 10^{19}$ eV are analyzed by using data of the Yakutsk and AGASA (Japan) extensive air shower (EAS) arrays. It is supposed that the clusters can be formed as a result of decay of superheavy nuclei. The consequences of this supposition compare with experimental data.

## 1. Introduction

The origin of cosmic rays is one of fundamental problems in the high-energy astrophysics. The origin of clusters in the cosmic rays is the unsolved question. At present it is unknown how the clusters appear. In the work [1] it is assumed that the showers in clusters are formed by neutral particles which arrive from cosmic ray sources. In [2] it is proposed that the clusters are formed as a result of focusing of charged particles by the galactic magnetic field. Some authors consider that the clusters are formed by accident [3].

## 2. Experiment

At the Yakutsk EAS array for period 30 years 29 showers with $E > 4 \cdot 10^{19}$ eV have been detected whose axes are inside of the array perimeter. Among them we found 2 clusters (doublets) [4] with the distance between the showers less than $5°$ (Fig.1, triangles, Y1, Y2). The distribution of clusters is presented on the equiexposure map of the celestial sphere in the 2nd equatorial coordinate system ($\delta$ - declination, RA – right ascension). On such a map (as compared with the usual astronomical map the scale of declination is changed) it is expected the same number of showers from equal areas. The probability of random formation of these clusters is 0.1.

Earlier, at relatively low energies, $E \sim 1 \cdot 10^{19}$ eV, we found 4 clusters by using data of world EAS arrays [5]. The probability of chance was $P < 2.7 \times 10^{-3}$. These clusters were distributed isotropy.

At the AGASA array 57 showers with $E > 4 \cdot 10^{19}$ eV have been registered [1] among which there are 6 clusters: 1 triplet and 5 doublets (Fig1, circles, A1,…, A6). In these clusters the distance between showers is $< 2.5°$. The chance probability to form 6 clusters from 57 showers is $\sim 10^{-2}$ [1]. As seen from Fig.1, the clusters are distributed over the celestial sphere practically uniformly. Note that ~23% of the total number of showers form the clusters.

## 3. Decay of superheavy nuclei

In [6] we assumed that $E > 4 \cdot 10^{19}$ eV cosmic rays were, most likely superheavy nuclei with a charge $Z > 26$. In the framework of this supposition, the origin of clusters can be explained: they are formed as a result of the spontaneous decay of superheavy nuclei into individual second nuclei.

The consequences of this supposition are below:
1. The difference in arrival time of nuclei forming a cluster will depend on the summary energy of a primary nuclei. Than more energy the nuclei will propagate further from a source due to relativistic effect and there is a high probability, that nuclei will decay nearer to the Earth. For second nuclei that have originated near to the Earth, the difference in arrival time will be minimal.

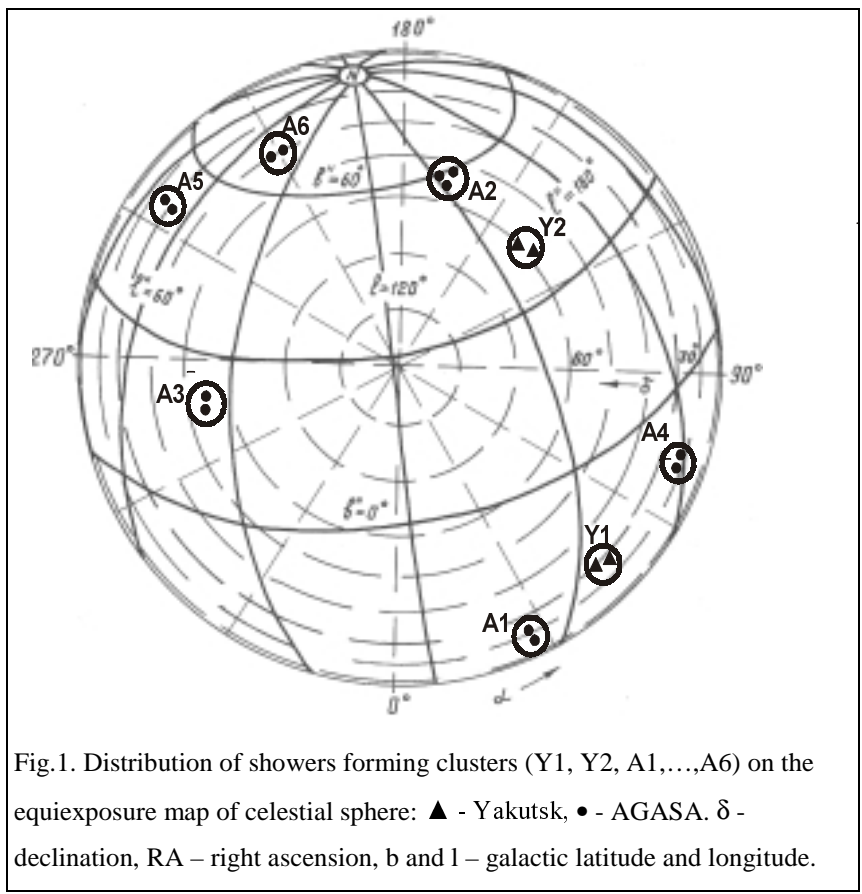

Fig.1. Distribution of showers forming clusters (Y1, Y2, A1,…,A6) on the equiexposure map of celestial sphere: ▲ - Yakutsk, ● - AGASA. δ - declination, RA – right ascension, b and l – galactic latitude and longitude.

2. Because of increase the probability of nucleus decay with energy the increase of cluster number is possible.

3. The mass composition of second nuclei will be more light in comparison of the primary nuclei. The showers generated by these second nuclei will differ from showers of primary nuclei.

## 4. Discussion

Consider below whether the above predictions are realized for experimental data of the Yakutsk and AGASA arrays. The Table presents the arrival time, energy, coordinates of showers forming clusters according to Yakutsk and AGASA array data [1].

1. Fig.2 shows the summary energy of showers of a cluster and difference in arrival time of these showers. As seen from Fig.2, for the lower the summary energy of showers in a cluster it corresponds the larger the difference in time of their arrival. It can be assumed that the summary energy of nuclei in the clusters is proportional to the primary nuclei energy. Thus, the first condition is confirmed.

Table. Clusters by Yakutsk (Y1, Y2) and AGASA (A1, …A6) array data.

| Clusters | Date | Energy, $10^{19}$ eV | δ, grad. | RA, grad. |
|---|---|---|---|---|
| Y1 | 1978.10.06 | 4.2 | 24.9 | 47.9 |
|    | 1980.05.16 | 4.3 | 29.2 | 46.9 |
| Y2 | 1992.05.02 | 6.7 | 60.7 | 131.2 |
|    | 2001.10.31 | 4.0 | 59.6 | 128.0 |
| A1 | 1993.12.03 | 21.3 | 21.1 | 18.75 |
|    | 1995.10.29 | 5.07 | 20.0 | 18.5 |
| A2 | 1992.08.01 | 5.50 | 57.1 | 172.25 |
|    | 1995.01.26 | 7.76 | 57.6 | 168.5 |
|    | 1998.04.04 | 5.35 | 56.0 | 168.25 |
| A3 | 1991.04.20 | 4.35 | 47.8 | 284.75 |
|    | 1994.07.06 | 13.4 | 48.3 | 281.25 |
| A4 | 1986.01.05 | 5.47 | 30.1 | 69.5 |
|    | 1995.11.15 | 4.89 | 29.9 | 70.25 |
| A5 | 1996.01.11 | 14.4 | 23.0 | 241.5 |
|    | 1997.04.10 | 3.89 | 23.7 | 239.5 |
| A6 | 1996.12.24 | 4.97 | 37.7 | 214.25 |
|    | 2000.05.26 | 4.98 | 37.1 | 212.0 |

2. According to AGASA data, a portion of the total number of showers forming clusters [1] increases with the energy.

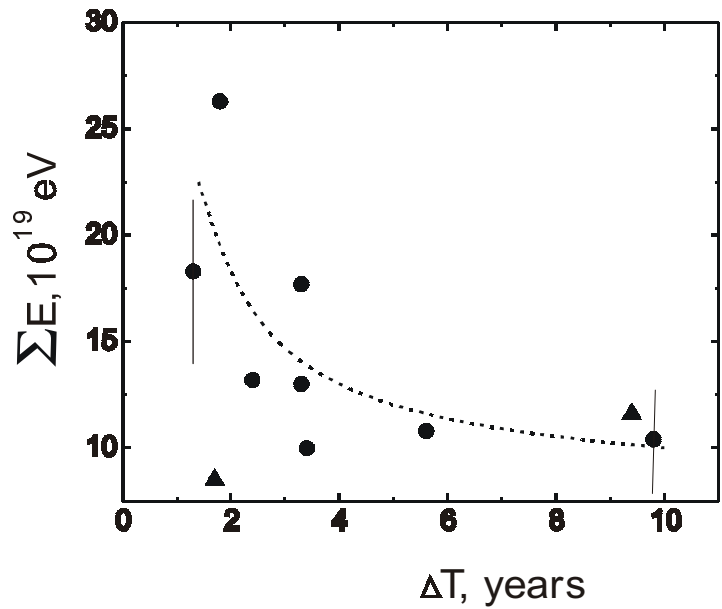

Fig.2. Summary energy of showers $\Sigma E$ forming a cluster and their difference in time of arrival $\Delta T$. The curve is the approximation of AGASA data. ▲ - Yakutsk, ● - AGASA.

3. No information concerning to the difference of showers inside and outside clusters by Yakutsk and AGASA data.

At lower energies, E~$10^{19}$ eV, at the Yakutsk array the showers with poor muons had been founded. About the one-half of these showers form clusters [7].

It is a possible, further observations at E>4·$10^{19}$ eV will show whether the showers inside and outside the clusters differ in content of muons.

Thus, all predictions in the framework of the above supposition most likely are on the whole confirmed.

## 5. Conclusion

In conclusion it may be said that the clusters can be formed as a result of decay of superheavy nuclei.

The Yakutsk EAS array is supported by the Ministry of Education of the Russian Federation (a project no. 01-30).